\documentclass[aps,prx,reprint,amsmath,amssymb,floatfix,longbibliography,showkeys,preprintnumbers,superscriptaddress]{revtex4-2}
\usepackage[T1]{fontenc}
\usepackage[utf8]{inputenc}
\usepackage{graphicx}
\usepackage{bm}
\usepackage{booktabs,tabularx,array}
\usepackage{capt-of}
\usepackage{microtype}
\begin{document}
\title{Pulse-Level Compilation of Measurement-Free Recovery in Transmon Circuits}
\author{Yi-Han Yu}
\affiliation{Beijing National Laboratory for Condensed Matter Physics, Institute of Physics, Chinese Academy of Sciences, Beijing 100190, China}
\affiliation{School of Physical Sciences, University of Chinese Academy of Sciences, Beijing 100049, China}

\author{Kai Xu}
\email{kaixu@iphy.ac.cn}
\affiliation{Beijing National Laboratory for Condensed Matter Physics, Institute of Physics, Chinese Academy of Sciences, Beijing 100190, China}
\affiliation{School of Physical Sciences, University of Chinese Academy of Sciences, Beijing 100049, China}
\affiliation{Beijing Key Laboratory of Fault-Tolerant Quantum Computing, Beijing Academy of Quantum Information Sciences, Beijing 100193, China}
\affiliation{Hefei National Laboratory, Hefei 230088, China}
\affiliation{Beijing Key Laboratory for Advanced Quantum Technology, Beijing 100190, China}

\author{Heng Fan}
\email{hfan@iphy.ac.cn}
\affiliation{Beijing National Laboratory for Condensed Matter Physics, Institute of Physics, Chinese Academy of Sciences, Beijing 100190, China}
\affiliation{School of Physical Sciences, University of Chinese Academy of Sciences, Beijing 100049, China}
\affiliation{Beijing Key Laboratory of Fault-Tolerant Quantum Computing, Beijing Academy of Quantum Information Sciences, Beijing 100193, China}
\affiliation{Hefei National Laboratory, Hefei 230088, China}
\affiliation{Beijing Key Laboratory for Advanced Quantum Technology, Beijing 100190, China}
\date{September 5, 2026}

\begin{abstract}
Quantum error correction commonly relies on syndrome measurement, decoding, and conditional feedback. We numerically show that the conditional spectrum of an interacting transmon circuit can compile a local recovery rule into a fixed open-loop control cycle. In a four-bit repetition-code ring, the resulting input-independent control slows the decay of logical coherence under Pauli-\(X\) noise and stabilizes logical-one population under data relaxation relative to uncorrected references. Its local, bounded-degree architecture provides a hardware-native framework for extending compiled recovery control to larger quantum networks.
\end{abstract}
\maketitle

\section{Introduction}

Large-scale universal quantum computation will require quantum error correction (QEC) \cite{L026,L027,L028,L062} to operate repeatedly \cite{L030,L031,L034,L036} with a control architecture that does not let syndrome processing and recovery dominate the cycle overhead \cite{L004,L017,L032,L033}. Conventional cycles measure syndromes, decode them, and apply measurement-conditioned feedback \cite{L006,L041,L045,L050}, placing demanding requirements on readout fidelity, measurement time, feedback latency, and control routing. Measurement-free QEC removes the intermediate readout \cite{L001,L002}, but it can reorganize these control tasks in different ways. Coherent implementations \cite{L003,L004,L016,L017} can retain explicit syndrome-dependent circuit operations \cite{L005,L013,L018}, whereas autonomous approaches \cite{L044,L046,L047,L048} embed recovery in engineered drive-and-dissipation dynamics \cite{L007,L008,L009,L010} or reset-mediated reservoirs \cite{L011} tailored to a chosen encoding and protected manifold \cite{L049}. Here we explore a complementary route: compiling joint-syndrome selection and recovery into a fixed open-loop waveform for each local update.

Figure~1 gives the physical picture: spectral selectivity lets a fixed drive on a leaf auxiliary respond only to a local mismatch, transferring its error record out of the encoded data network. Reset then prepares the auxiliary for reuse, so that syndrome selection and recovery are compiled into the local dynamics rather than enacted by online feedback. We support this picture with numerical simulations, first locally, then on a four-site ring driven by one input-independent waveform, and finally under repeated cycles for logical coherence under Pauli-\(X\) noise and logical-one population under data relaxation. We further examine the feasibility and limits of extending this approach to larger networks.
\begin{figure}[!t]
\centering
\includegraphics[width=0.98\columnwidth,keepaspectratio]{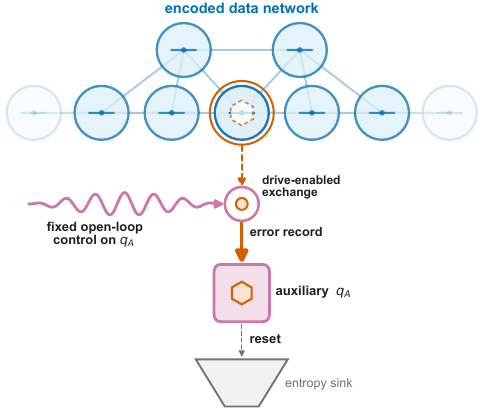}
\caption{\textbf{Fixed open-loop control transfers a local error record to a leaf auxiliary.} The selected data neighborhood and preset composite waveform meet at the control junction, after which the record resides in \(q_A\). The dashed outlet denotes ideal auxiliary re-preparation between uses of \(q_A\); it is neither measurement-triggered feedback nor a simulated finite-duration reset.}
\label{fig1}
\end{figure}

\section{Spectrally resolved compilation of a local recovery predicate}

\subsection{Target operation and minimal spectral compiler}

We consider the four-mode local module in Fig.~2(c), containing three data modes, \(q_L\), \(q_M\), and \(q_R\), and one auxiliary mode, \(q_A\). It updates \(q_M\) only when \(q_M\) disagrees with two mutually consistent neighbors.

For the repetition code, the local checks are \(S_L=Z_LZ_M\) and \(S_R=Z_MZ_R\). With \(Z|b\rangle=(-1)^b|b\rangle\), a middle-bit error has joint syndrome \((S_L,\allowbreak S_R)=(-1,\allowbreak -1)\). The corresponding correction sector is

\[
\Pi_M^{\mathrm{err}}
=\frac{1}{4}(I-S_L)(I-S_R).
\]

which has support only on \(|010\rangle\) and \(|101\rangle\). With abstract ready and record states, the intended operation is

\[
V_M|\psi\rangle_D|r\rangle_A
=
(I-\Pi_M^{\mathrm{err}})|\psi\rangle_D|r\rangle_A
+
X_M\Pi_M^{\mathrm{err}}|\psi\rangle_D|e\rangle_A .
\]

At the selected working point, \(|r\rangle_A\) and \(|e\rangle_A\) correspond to the \(a1\) and \(a0\) states of \(q_A\). The complete waveform must therefore map \((010,a1)\rightarrow(000,a0)\) and \((101,a1)\rightarrow(111,a0)\).

The projector therefore specifies a local middle-target primitive: for a fixed \(a1\) initialization, it maps \(010\rightarrow000\) and \(101\rightarrow111\), while leaving the other six computational-basis states unchanged. The two corrected branches carry the common auxiliary assignment \(a0\). Sec.~III applies the same local rule at every target site of a ring.

The static \(ZZ\) couplings make the \(q_M\) transition frequencies depend on the states of \(q_L\) and \(q_R\). We use these conditional shifts as a control resource: frequency-selective drives address the branches selected by \(\Pi_M^{\mathrm{err}}\) while leaving the other branches inactive \cite{L059,L060}.

The compiled recovery has two ordered steps. First, a photon-number-conserving swap addresses

\begin{center}
\resizebox{0.97\columnwidth}{!}{$\displaystyle
\mathcal T_M^{\mathrm{rec}}
=
\left\{
(0,1,0,1)\leftrightarrow(0,0,0,2),\,
(1,0,1,1)\leftrightarrow(1,1,1,0)
\right\},
$}
\end{center}

where each tuple lists the occupations of \((q_L,\allowbreak q_M,\allowbreak q_R,\allowbreak q_A)\). The two selected inputs are transferred to \((000,a2)\) and \((111,a0)\) by swaps in opposite directions.

Second, a photon-number-nonconserving conditional auxiliary \(0\leftrightarrow2\) transition addresses

\[
\mathcal T_M^{02}
=
\left\{
(0,0,0,2)\leftrightarrow(0,0,0,0)
\right\},
\]

and maps \((000,a2)\) to \((000,a0)\). The \(0\leftrightarrow2\) step is a two-photon activation of a nominally forbidden transition. The \(0\leftrightarrow2\) transition has been demonstrated in multiple works \cite{L063,L064,L065}. The \((111,a0)\) branch is already in the final auxiliary level and is not driven. Section~II~B gives the physical implementation of both steps and tests the complete local waveform.

\begin{figure*}[!t]
\centering
\includegraphics[width=0.96\textwidth]{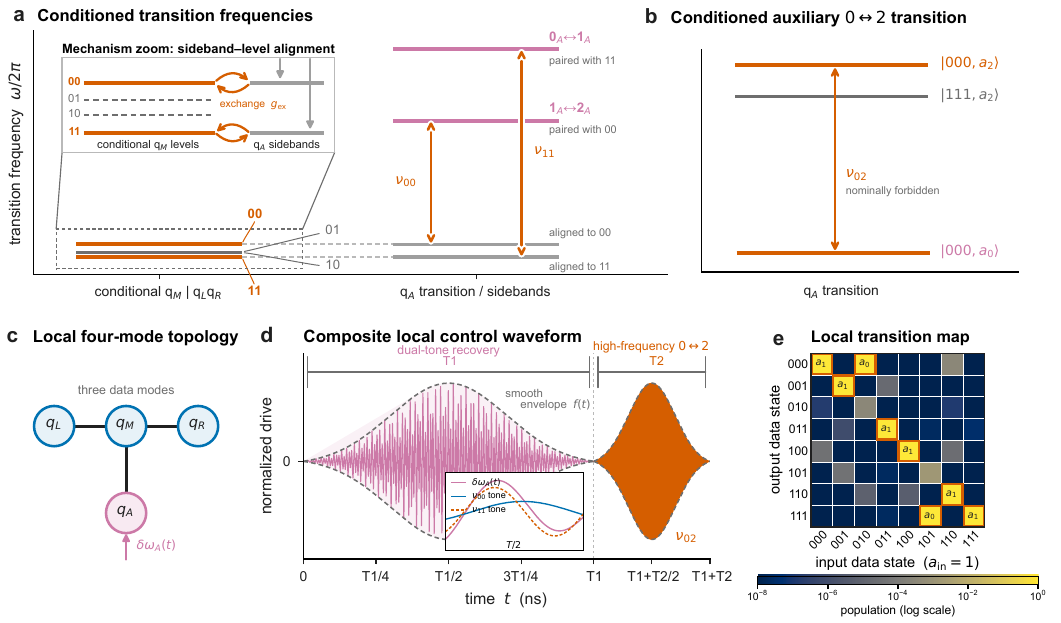}
\caption{\textbf{Condition-selective local compilation in a four-mode module.} \textbf{a,} Conditioned transition frequencies and modulation-generated sidebands at the selected working point. \textbf{b,} Conditioned auxiliary \(0\leftrightarrow2\) transition: the \(00\)-conditioned line is selected while the \(11\) branch remains a spectator. \textbf{c,} Local topology: \(q_L-q_M-q_R\) form the data chain, while leaf auxiliary \(q_A\) couples to \(q_M\) and receives \(\delta\omega_A(t)\). \textbf{d,} Composite local control waveform, consisting of the dual-tone recovery segment followed by the conditional auxiliary \(0\leftrightarrow2\) segment. \textbf{e,} Assigned-data populations for the complete local waveform over all eight inputs. Cell color sums over auxiliary occupations; highlighted cells mark the prescribed target data labels, and the in-cell \(a0/a1\) labels specify the corresponding target auxiliary assignment. Population outside the displayed data subspace is retained without renormalization.}
\label{fig2}
\end{figure*}

\begin{figure*}[!t]
\centering
\includegraphics[width=0.96\textwidth]{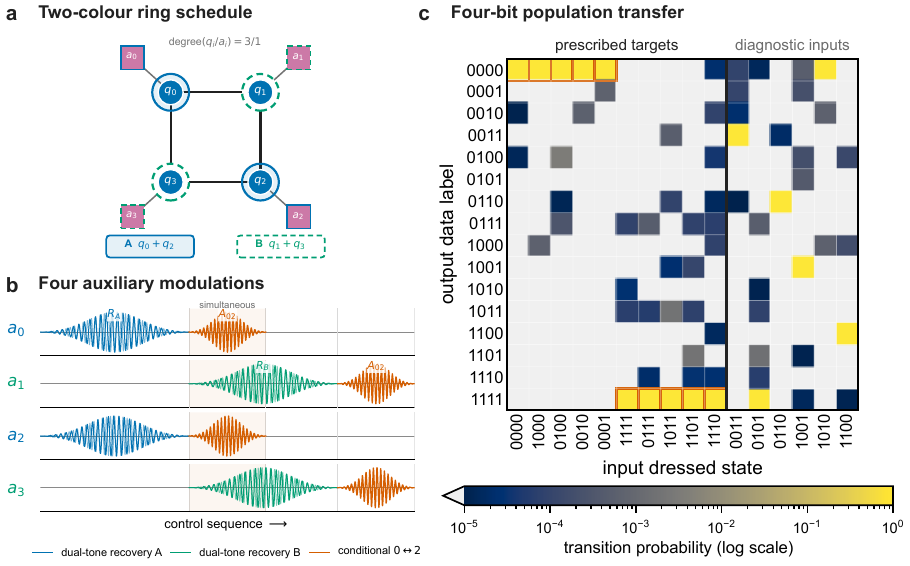}
\caption{\textbf{Pipelined ring schedule and four-bit population transfer.} \textbf{a,} The four-bit ring contains four data modes and one leaf auxiliary per data mode; the recovery rule groups nonadjacent targets as \(A=\{q_0,\allowbreak q_2\}\) and \(B=\{q_1,\allowbreak q_3\}\). The dynamical model retains four levels for all eight modes. \textbf{b,} The fixed input-independent waveform pipelines the two recovery groups with their conditional auxiliary \(0\leftrightarrow2\) subpulses. Each auxiliary operation begins after the corresponding recovery operation, while nonconflicting operations overlap. \textbf{c,} Assigned-data populations after the composite waveform, with columns denoting inputs and rows output labels. All 16 inputs are propagated, but only the two codewords and eight single-flip inputs have prescribed full-system dressed targets; the other six columns are double-flip diagnostics. The matrix reports population transfer, not a coherent process.}
\label{fig3}
\end{figure*}

\subsection{Physical realization of the composite local waveform}

The local module realizes the compiler through two spectrally selected processes. In the circuit of Fig.~2(c), fixed capacitive couplings between \(q_L-q_M-q_R\) and between \(q_M\) and the leaf auxiliary \(q_A\) generate conditional shifts. At the selected working point, these shifts resolve the \(00\)- and \(11\)-conditioned data--auxiliary exchange lines from the \(01\) and \(10\) spectator sectors [Fig.~2(a)]. Starting with \(q_A\) in \(\lvert1\rangle_A\), we apply the longitudinal two-tone modulation

\[
\frac{H_{\mathrm{rec}}(t)}{h}
=
f_{\mathrm{rec}}(t)\,\hat n_A
\sum_{\eta\in\{00,11\}}
A_\eta\cos(2\pi\nu_\eta t).
\]

Its two carriers address \(\lvert\widetilde{010,1}\rangle\leftrightarrow\lvert\widetilde{000,2}\rangle\) and \(\lvert\widetilde{101,1}\rangle\leftrightarrow\lvert\widetilde{111,0}\rangle\). The resulting opposite-direction exchanges correct the middle data mode while placing the two selected branches in \((000,a2)\) and \((111,a0)\).

The recovery exchange is followed by a second, independently conditioned process. The auxiliary \(0\leftrightarrow2\) frequency is also shifted by the data configuration [Fig.~2(b)]. We therefore drive the \(000\)-conditioned line with the photon-number-nonconserving auxiliary modulation

\[
\frac{H_{02}(t)}{h}
=
f_{02}(t)
\left(a_A^2+a_A^{\dagger2}\right)
A_{02}\cos(2\pi\nu_{02}t+\phi_{02}).
\]

Applied after the recovery segment, this pulse transfers \((000,a2)\) to \((000,a0)\). The \((111,a0)\) branch and the six branches that remain in \(a1\) are spectrally off-resonant. Figure~2(d) shows the resulting fixed composite waveform: a dual-tone recovery segment followed by the conditional auxiliary \(0\leftrightarrow2\) segment.

The interval between the dual-tone recovery and conditional auxiliary \(0\leftrightarrow2\) pulses provides a continuous control parameter for tuning recovery-induced logical phase offsets through the differential dynamical phase accumulated by the intermediate dressed states.

We test this complete physical sequence in a four-level full-Hamiltonian calculation of the four active transmons. We propagate the same composite waveform from each of the eight local computational-basis inputs and evaluate the assigned output branches. Figure~2(e) shows the corresponding assigned-data populations: \(010,a1\) and \(101,a1\) reach \(000,a0\) and \(111,a0\), while the other six inputs retain their data labels and auxiliary assignment \(a1\). The residual error in these state-transition probabilities is dominated by leakage induced by spectral crowding; an appropriate frequency arrangement of neighboring qubits can mitigate this limitation. Appendix~A~1 distinguishes the recovery-subpulse and complete-waveform calculations and gives their respective waveform parameters and state assignments. Section~III deploys this validated local waveform in the four-site ring schedule.
\section{Closed four-site ring realization}

\subsection{Bounded-degree extension and two-color recovery schedule}

To extend the local compiler, consider an even ring of \(N\) data modes. At each target \(i\), the local predicate becomes
\[
\Pi_i^{\mathrm{err}}
=
\frac{(I-Z_{i-1}Z_i)(I-Z_iZ_{i+1})}{4},
\]
with one leaf auxiliary assigned to that target. Each update still involves only the target, its two neighbors, and its auxiliary, so the local decision unit and the data and auxiliary mode degrees remain fixed as the ring grows. The remaining question is how to schedule these local updates.

Neighboring updates cannot generally share a recovery window: modulation sidebands can perturb the conditionally resolved transition of an adjacent recovery \cite{L060,L061}. We therefore drive only nonadjacent targets together. Because an even ring is bipartite, its targets split into two nonadjacent color classes, giving a two-stage waveform schedule whose number of stages does not grow with \(N\).

The \(N=4\) ring in Fig.~3(a) is the smallest closed instance with two distinct neighbors per target and a nontrivial two-color partition, \(A=\{q_0,\allowbreak q_2\}\) and \(B=\{q_1,\allowbreak q_3\}\). Its fixed waveform applies recovery on \(A\), overlaps the corresponding auxiliary conditioning with recovery on \(B\), and then applies the \(B\)-group auxiliary subpulses [Fig.~3(b)]. Appendix~A~2 defines the waveform. Section~III~B implements this schedule in the four-site ring and evaluates its complete input population map.

\subsection{Numerical ring experiment and complete population map}

We test the fixed two-color waveform in an eight-mode, four-level ring model by propagating all 16 computational-basis data inputs with the auxiliaries initialized in \(a1\). Appendix~A~2 gives the ring model, waveform, and scoring definition. The \(A=\{q_0,q_2\}\) recovery subpulses run first; their conditional auxiliary \(0\leftrightarrow2\) subpulses overlap the recovery window for \(B=\{q_1,q_3\}\) because the selected conditional transitions do not conflict. Figure~3(c) shows the resulting \(16\times16\) assigned-data population map, with raw final populations summed over auxiliary occupations. For both codewords and all eight single-flip inputs, the prescribed target is the dominant output; \(0111\rightarrow1111\) is the least favorable scored branch.

The six weight-two inputs have no unique recovery target: each is equally distant from the two codewords, so the repetition encoding does not specify which codeword to restore. They are therefore diagnostics of the fixed schedule, not scored recovery cases. This limitation comes from the code's correctable error set rather than from conditional spectral control. Since Fig.~3(c) sums over auxiliary occupations, it reports data-population transfer, not a coherent process or a process matrix. After auxiliary reset, the same waveform can be repeated; Section~III~C examines that repeated-cycle setting.

\begin{figure}[!t]
\centering
\includegraphics[width=0.98\columnwidth,keepaspectratio]{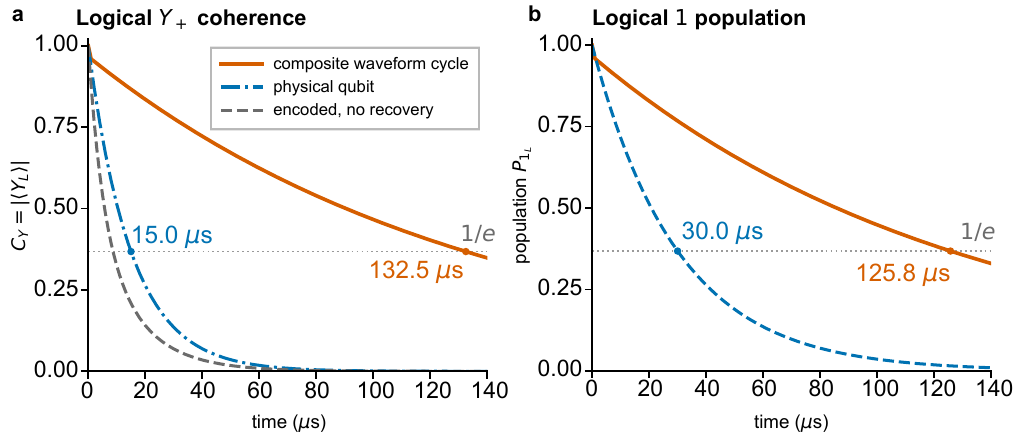}
\caption{\textbf{Repeated-cycle protection under two noise models.} \textbf{a,} Under dressed Pauli-\(X\) noise, recovery slows the decay of logical coherence \(C_Y\) relative to both a physical qubit and the unrecovered encoded state. \textbf{b,} Under independent amplitude damping, recovery slows the decay of logical-one population \(P_{1_L}\) relative to a physical excited state. Both panels use the same recovery cycle but different initial states and observables.}
\label{fig4}
\end{figure}

\subsection{Repeated-cycle protection and code-defined error scope}

Resetting the auxiliaries closes the fixed waveform into a repeated recovery cycle. The effective cycle model and the two noise channels are defined in Appendix~A~3. Under independent Pauli-\(X\) noise, we start from \(\lvert+i_L\rangle=(\lvert0000\rangle+i\lvert1111\rangle)/\sqrt{2}\) and monitor the logical coherence \(C_Y=|\langle Y_L\rangle|\). The recovery cycle slows the decay of \(C_Y\) relative to both one physical qubit and the encoded state without recovery [Fig. 4(a)].

Under independent amplitude damping, we instead start from the high-excitation codeword \(\lvert1_L\rangle=\lvert1111\rangle\) and track the logical-one population \(P_{1_L}\). The same recovery cycle slows the decay of \(P_{1_L}\) relative to one physical excited state [Fig. 4(b)]. The repetition encoding is not designed to correct \(T_1\) decay, so this calculation follows the stability of the populated logical branch through repeated cycles.

The error axis is fixed by the repetition encoding. With \(Z_iZ_{i+1}\) checks, a bit flip \(X_i\) reverses its neighboring syndromes and enters the one-flip sector addressed by the waveform, whereas a phase flip \(Z_i\) leaves the checks unchanged and lies outside the present recovery rule. Section~IV considers the corresponding phase-error route and its additional control resources.
\section{Generality and hardware constraints}

The four-site repetition-code result realizes one compiled recovery rule. More generally, let \(\{S_k\}\) be commuting stabilizers \cite{L029} with syndrome eigenvalues \(s_k\in\{-1,\allowbreak +1\}\). A Boolean predicate \(g(\mathbf{s})\) selects the syndrome sectors that require a recovery action through

\[
\Pi_g
=
\sum_{\mathbf{s}:g(\mathbf{s})=1}
\prod_k\frac{I+s_kS_k}{2}.
\]

For the local repetition code, \(g\) accepts only \((s_L,\allowbreak s_R)=(-1,\allowbreak -1)\), so \(\Pi_g\) reduces to \(\Pi_M^{\mathrm{err}}\). The predicate specifies the branches that the waveform must address. Its hardware realization requires every accepted branch to admit an ordered dressed-state path with resolvable transition frequencies, usable drive matrix elements, and a feasible schedule on the hardware graph.

Basis conjugation gives a corresponding route for the phase-error boundary in Sec.~III~C. Since \(HZH=X\), an \(X\)-basis repetition code presents phase flips as bit-flip syndromes in the rotated basis. Realizing that route would add basis rotations and phase-sensitive calibration to the compiler; the present calculations address only the \(Z\)-basis rule above.

Algebraic expressibility, spectral implementability, and low-resource realization are therefore distinct requirements. A practical realization further requires bounded connectivity, a manageable tone count, a shallow schedule, and controlled branch phases. The predicate-level formalism defines the recovery task, while the hardware graph and dressed spectrum determine the waveform that realizes it.
\section{Conclusion}

We have shown that a local quantum-error-correction rule can be compiled directly into a fixed open-loop control cycle, rather than assembled from syndrome measurement, classical decoding, and conditional feedback. In an interacting superconducting circuit, the conditional dressed spectrum provides a native physical resource for control: its state-dependent transitions encode the local syndrome predicate and allow the prescribed recovery path to be addressed within a fixed waveform. The resulting protocol combines syndrome selection and recovery within one hardware-native waveform, providing a pulse-level realization of measurement-free local correction.

For a transmon implementation of a four-bit repetition-code ring, we demonstrated that this compiled control can be applied across the ring with a bounded-degree local architecture and a parallel two-color schedule. The corresponding repeated-cycle calculations show protection of the logical observables appropriate to the two noise models considered. More broadly, the result suggests a compiler viewpoint for measurement-free QEC: the code specifies the recovery predicate, while the hardware spectrum, connectivity, and control resources determine how that predicate can be realized. This separation provides a route for translating other local recovery rules into platform-specific pulse programs and for extending measurement-free correction beyond gate-by-gate constructions.

\begin{acknowledgments}
We thank Zheng-Yi Han for helpful discussions. We acknowledge support from the Synergetic Extreme Condition User Facility (SECUF) in Huairou District, Beijing. This work was supported by the National Natural Science Foundation of China (Grants No. 92265207, No. T2121001, No. U25A6009, No. T2322030, No. 12122504, No. 12274142, and No. 12475017), the Innovation Program for Quantum Science and Technology (Grant No. 2021ZD0301800), the Beijing Nova Program (Grant No. 20220484121), and the Ministry of Science and Technology project (Grant No. 2025YFE0217600).
\end{acknowledgments}

\appendix
\section{Numerical models and observables}

In expressions for \(H/\hbar\), \(\omega\), \(\alpha\), \(g\), and \(\Omega\) denote angular frequencies. Carrier frequencies \(\nu\) and the amplitudes \(A\) in expressions for \(H/h\) are cyclic frequencies. Tables quote cyclic-frequency values in MHz, durations in ns, and phases in radians; thus a propagator uses \(2\pi\times10^{-3}(H/h)_{\mathrm{MHz}}t_{\mathrm{ns}}\). We distinguish a qutrit recovery subpulse, the four-level complete local waveform of Fig.~2(e), and the four-level eight-mode ring model of Fig.~3. Figure~4 uses the effective channel derived from the ring waveform.

\subsection{Local models for Fig.~2}

The local calculation contains three data transmons \(q_L,q_M,q_R\) and one auxiliary \(q_A\). With \(b_j=\sum_{m=0}^{1}\sqrt{m+1}\lvert m_j\rangle\langle m+1_j\rvert\), \(n_j=b_j^\dagger b_j\), and charge operators \(Q_j\) projected into the retained local eigenbasis and normalized by \(|\langle0_j|\hat n_j^{\mathrm{ch}}|1_j\rangle|\), the qutrit recovery model is
\[
\begin{aligned}
\frac{H_0^{\mathrm{loc}}}{\hbar}={}&
\sum_{j\in\{L,M,R,A\}}\left(\omega_jb_j^\dagger b_j-\frac{\alpha_j}{2}b_j^{\dagger2}b_j^2\right)
\\ &+g_{LM}Q_LQ_M+g_{MR}Q_MQ_R+g_{MA}Q_MQ_A .
\end{aligned}
\tag{A1}
\]
Table~\ref{tab:local-parameters} records the two local calculations and their pulse parameters.  The complete four-level model uses the fitted cosine-transmon eigenenergies for the local terms, with the full projected charge matrices for the same three links. Its transformed charge couplings retain counter-rotating terms; the recovery segment is followed by the conditional segment.

\begin{center}
\squeezetable
\resizebox{\columnwidth}{!}{%
\begin{tabular}{lcc}
\toprule
& \textbf{qutrit recovery} & \textbf{Fig.~2(e) complete local} \\
\midrule
active dimensions & \(3^4\) & \(4^4\) \\
\(\omega_{L,M,R,A}/2\pi\) & \((5375,5750,5275,6300)\) & \((5350,5750,5250,6300)\) \\
\(\alpha_{L,M,R,A}/2\pi\) & \((250,250,250,200)\) & \((250,250,250,200)\) \\
\((g_{LM},g_{MR},g_{MA})/2\pi\) & \((60,60,25)\) & \((60,60,60)\) \\
charge coupling & projected \(Q_jQ_k\) & full transformed charge \\
recovery \((T,\sigma,\varphi)\) & \((200,40,0)\) & \((200,40,0)\) \\
\((\nu_{00},\Omega_{00}/2\pi)\) & \((339.604136,24.907372)\) & \((364.932886,11.121676)\) \\
\((\nu_{11},\Omega_{11}/2\pi)\) & \((571.668736,58.368773)\) & \((599.951033,29.259976)\) \\
conditional \((T_{02},\sigma_{02})\) & --- & \((80,16)\) \\
\((\nu_{02},A_{02},\phi_{02})\) & --- & \((12419.020565,9.623758,1.230795)\) \\
\bottomrule
\end{tabular}
}
\captionof{table}{\label{tab:local-parameters}Local-model parameters. Frequencies, anharmonicities, couplings, and amplitudes are in MHz; durations are in ns.}
\end{center}

The recovery subpulse is
\[
\frac{H_{\mathrm{rec}}^{\mathrm{loc}}(t)}{\hbar}=
\frac{H_0^{\mathrm{loc}}}{\hbar}+G_{T,\sigma}(t)\sum_{\mu\in\{00,11\}}\Omega_\mu\cos(2\pi\nu_\mu t+\varphi_\mu)n_A ,
\tag{A2}
\]
where \(G_{T,\sigma}\) is a zero-edge, area-normalized Gaussian. Product labels are assigned to static dressed states by maximum bare-label overlap, giving
\[
\begin{aligned}
\lvert\widetilde{010,a1}\rangle&\rightarrow\lvert\widetilde{000,a2}\rangle,\\
\lvert\widetilde{101,a1}\rangle&\rightarrow\lvert\widetilde{111,a0}\rangle,\\
\lvert\widetilde{\mathbf b,a1}\rangle&\rightarrow\lvert\widetilde{\mathbf b,a1}\rangle,\\
\mathbf b&\in\{000,001,011,100,110,111\}.
\end{aligned}
\tag{A3}
\]

The conditional segment of the Fig.~2(e) calculation is
\[
\frac{H_{02}^{\mathrm{loc}}(t)}{h}=F_{T_{02},\sigma_{02}}(t)A_{02}\cos(2\pi\nu_{02}t+\phi_{02})\left(a_A^2+a_A^{\dagger2}\right),
\tag{A4}
\]
where \(a_A\) is the four-level auxiliary lowering operator and \(F_{T_{02},\sigma_{02}}\) is a zero-edge, peak-normalized Gaussian. The parameters are listed in Table~\ref{tab:local-parameters}. The final assigned branches are \(010,a1\rightarrow000,a0\), \(101,a1\rightarrow111,a0\), and \(\mathbf b,a1\rightarrow\mathbf b,a1\) for the remaining six strings. For Fig.~2(e), each product label \((\mathbf d,a)\) is assigned to the static eigenstate \(\lvert\widetilde{\mathbf d,a}\rangle\) with maximum overlap. The displayed cell is \(\sum_{a=0}^{3}|\langle\widetilde{\mathbf d,a}|\psi_x(T)\rangle|^2\), whereas the prescribed full-branch probability contains only the specified auxiliary label. Population outside these assigned states is not renormalized into the displayed map.

\subsection{Four-level ring model for Fig.~3}

The ring has four data modes \(a_i\), four leaf auxiliaries \(b_i\), and edges \(\mathcal E=\{(0,1),(1,2),(2,3),(3,0)\}\). Its static Hamiltonian before the RWA is
\begin{equation}
\begin{aligned}
\frac{H_0^{(4)}}{\hbar}={}&\sum_{i=0}^{3}\left[\omega_i a_i^\dagger a_i+\frac{\alpha_i}{2}a_i^{\dagger2}a_i^2\right.\\
&\left.\qquad+\omega_{A_i}b_i^\dagger b_i+\frac{\alpha_{A_i}}{2}b_i^{\dagger2}b_i^2\right]\\
&+\sum_{i=0}^{3}g_i(a_i^\dagger b_i+a_i b_i^\dagger)\\
&+\sum_{(i,j)\in\mathcal E}\left\{-\frac{E_{Jc,ij}}{\hbar}\left[\cos(\phi_i-\phi_j)\right.\right.\\
&\left.\left.\qquad-\cos\phi_i-\cos\phi_j+I\right]+\frac{J_{ij}}{\hbar}q_iq_j\right\}.
\end{aligned}
\tag{A5}
\end{equation}
Here \(q_i=i(a_i^\dagger-a_i)\) and \(\phi_i=\phi_{i,\mathrm{zpf}}(a_i+a_i^\dagger)\). The phase scale is \(\phi_{i,\mathrm{zpf}}=(2E_{C,i}/E_{J,i})^{1/4}\), with \(E_{C,i}/h=|\alpha_i|/(2\pi)\) and \(E_{J,i}/h=[\omega_i/(2\pi)+E_{C,i}/h]^2/[8E_{C,i}/h]\). The cosine subtraction avoids double counting local terms already represented by the Duffing Hamiltonian.

To suppress frequency-crowding-induced hybridization in the data ring, each capacitive data--data link is paralleled by a SQUID-mediated Josephson channel. The capacitive term is set so that destructive interference nulls the computational single-excitation exchange \(\langle10|V_{ij}|01\rangle\), while the nonlinear SQUID term retains a finite \(ZZ\) interaction.

All modes retain four levels. The static parameters are given in Table~\ref{tab:ring-static}.

\begin{center}
\squeezetable
\resizebox{\columnwidth}{!}{%
\begin{tabular}{lr}
\toprule
\textbf{Eight-mode parameter} & \textbf{Value} \\
\midrule
\((\omega_0,\omega_1,\omega_2,\omega_3)/2\pi\) & \((5580,4980,5520,4820)\) \\
\((\omega_{A_0},\omega_{A_1},\omega_{A_2},\omega_{A_3})/2\pi\) & \((6110,5500,6040,5350)\) \\
\(\alpha_i/2\pi=\alpha_{A_i}/2\pi\) & \(-250\) \\
\((E_{Jc,01},E_{Jc,12},E_{Jc,23},E_{Jc,30})/h\) & \((655.330726,652.530968,644.411161,647.176081)\) \\
\((J_{01},J_{12},J_{23},J_{30})/h\) & \((99,99,99,99)\) \\
\((g_0,g_1,g_2,g_3)/2\pi\) & \((29,28,28,29)\) \\
\(\omega_{\mathrm{rot}}/2\pi\) & \(5500\) \\
\bottomrule
\end{tabular}
}
\captionof{table}{\label{tab:ring-static}Static ring parameters, in MHz.}
\end{center}

With \(\hat N=\sum_i(a_i^\dagger a_i+b_i^\dagger b_i)\), the static RWA is
\[
H_0^{\mathrm{RWA}}=\sum_NP_NH_0^{(4)}P_N,\qquad H_{\mathrm{rot}}=H_0^{\mathrm{RWA}}-\hbar\omega_{\mathrm{rot}}\hat N .
\tag{A6}
\]
Here \(P_N\) projects onto total excitation number \(N\). The recovery modulation is \(\delta\omega_{A_i}(t)=\sum_pG_p(t)\Omega_p\cos(2\pi\nu_pt+\varphi_p)\), where the sum includes the active recovery tones. The drives are
\[
\begin{aligned}
\frac{V_{\mathrm{rec}}(t)}{\hbar}&=\sum_i\delta\omega_{A_i}(t)n_{A_i},\\
\frac{V_{02}(t)}{\hbar}&=\sum_i\sum_{p\in\mathcal P_i^{02}}\frac{G_p(t)\Omega_p}{2}\\
&\qquad\times\left(e^{-i\delta_pt}b_i^{\dagger2}+e^{i\delta_pt}b_i^2\right),\\
\delta_p&=2\pi\nu_p-2\omega_{\mathrm{rot}}.
\end{aligned}
\tag{A7}
\]
Recovery subpulses have \(T=200~\mathrm{ns}\); both recovery and conditional subpulses use zero-edge, area-normalized Gaussian envelopes with \(\sigma=0.2T\). The set \(\mathcal P_i^{02}\) contains the conditional tones on auxiliary \(i\), with envelopes zero outside their windows. All phases are zero. Table~\ref{tab:ring-pulses} lists the carrier and amplitude pairs; frequencies and amplitudes are in MHz and \(T_{02}\) is in ns.

\begin{center}
\squeezetable
\resizebox{\columnwidth}{!}{%
\begin{tabular}{lrrrrrrr}
\toprule
\(i\) & \(\nu_{00}\) & \(\Omega_{00}/2\pi\) & \(\nu_{11}\) & \(\Omega_{11}/2\pi\) & \(\nu_{02}\) & \(\Omega_{02}/2\pi\) & \(T_{02}\) \\
\midrule
0 & 308.033578 & 18.768230 & 587.300022 & 50.916564 & 11975.549786 & 4.459753 & 80 \\
1 & 297.553225 & 18.790512 & 577.240745 & 51.709313 & 10755.356348 & 2.229894 & 160 \\
2 & 297.857519 & 18.794686 & 577.792840 & 51.787347 & 11835.350606 & 4.459634 & 80 \\
3 & 308.121417 & 18.775307 & 587.822034 & 50.851408 & 10455.548306 & 4.459733 & 80 \\
\bottomrule
\end{tabular}
}
\captionof{table}{\label{tab:ring-pulses}Ring waveform parameters.}
\end{center}

With \(A=\{0,2\}\) and \(B=\{1,3\}\), the schedule is
\[
\begin{aligned}
R_A(0\text{--}200)&\rightarrow[A_A(200\text{--}280)\parallel R_B(200\text{--}400)]\\
&\rightarrow[A_{B,1}(400\text{--}560)\parallel A_{B,3}(400\text{--}480)].
\end{aligned}
\]

All 16 data strings start in their assigned dressed state with auxiliaries in \(a1111\). The two codewords and eight weight-one or weight-three strings have prescribed dressed targets; the six weight-two strings do not. For \(x\in\mathcal S_{10}\),
\[
P_{x,\mathrm{ring}}^{\mathrm{tar}}=\left|\langle\widetilde f_x|U_{\mathrm{comp}}|\widetilde i_x\rangle\right|^2,\qquad \overline P_{\mathrm{ring}}^{\mathrm{tar}}=\frac1{10}\sum_{x\in\mathcal S_{10}}P_{x,\mathrm{ring}}^{\mathrm{tar}}.
\tag{A8}
\]
Figure~3(c) uses a different label-grouping prescription. Let \(\lvert E_{N,k}\rangle\) be a static RWA eigenstate in excitation sector \(N\), and let \(\mathbf d_{N,k}\) be the data part of its largest bare product-state component. The displayed population in output row \(\mathbf d\), input column \(x\), is
\[
P_{\mathbf d\leftarrow x}^{\mathrm{assign}}
=\sum_{N=0}^{8}\sum_{k:\,\mathbf d_{N,k}=\mathbf d}
|\langle E_{N,k}|\psi_x(T)\rangle|^2 .
\tag{A9}
\]
States assigned noncomputational data labels and population in sectors above \(N=8\) are omitted from the displayed matrix without renormalization. This dressed-label population map differs from both the single-target probability in Eq.~(A8) and the bare product-basis marginal \(\sum_{\mathbf a}|\langle\mathbf d,\mathbf a|\psi_x(T)\rangle|^2\).

\subsection{Repeated-cycle model for Fig.~4}

The repeated-cycle model has the 16 dressed reset-basis labels \(\{\lvert E_{\mathbf d,a1111}\rangle\}\) plus a leakage sink. The waveform followed by an ideal phase-calibrated dressed reset defines
\[
\mathcal C_{\mathrm{comp}}(\rho)=\sum_\ell K_\ell\rho K_\ell^\dagger,\qquad\sum_\ell K_\ell^\dagger K_\ell=I_{17}.
\tag{A10}
\]
The Kraus construction pairs the two no-error codeword targets in one coherence group and the two logical partners of each single-bit syndrome in four further groups. Each selected target amplitude is phase calibrated to be positive real; the positive residual complement is sent to the leakage sink. The pulse durations, together with the inter-pulse interval discussed in Sec.~II~B, provide control parameters for compensating recovery-induced branch phases. The phase-calibrated channel used here represents the ideal outcome of such calibration. This effective construction assumes coherent dressed-state re-preparation and calibrated branch phases, without simulating their microscopic implementation.

Each \(1~\mu\mathrm{s}\) cycle contains the \(560~\mathrm{ns}\) waveform, ideal reset, and \(440~\mathrm{ns}\) idle interval:
\[
\mathcal M_{\mathrm{cyc}}=\mathcal Z_\phi\circ\mathcal N_{280\mathrm{ns}}\circ\mathcal C_{\mathrm{comp}}\circ\mathcal N_{280\mathrm{ns}}\circ\mathcal N_{440\mathrm{ns}} .
\tag{A11}
\]
The map \(\mathcal N_\tau\) applies the selected data-noise model for duration \(\tau\), and \(\mathcal Z_\phi\) removes the remaining deterministic logical phase. The two half-duration noise factors approximate noise during the waveform; they do not represent simultaneous driven open-system integration. For Fig.~4(a),
\[
\begin{aligned}
L_i&=\sqrt{\gamma_X}X_i^{(\mathrm{dr})},\qquad \gamma_X=(30~\mu\mathrm{s})^{-1},\\
X_i^{(\mathrm{dr})}&=\sum_{\mathbf d}\lvert E_{\mathbf d\oplus\mathbf e_i,a1111}\rangle\langle E_{\mathbf d,a1111}\rvert .
\end{aligned}
\tag{A12}
\]
The initial state and measured coherence are
\[
\begin{aligned}
\lvert+i_L\rangle&=\frac{\lvert E_{0000,a1111}\rangle+i\lvert E_{1111,a1111}\rangle}{\sqrt2},\\
C_Y(t)&=|\langle Y_L\rangle|,
\end{aligned}
\tag{A13}
\]
where \(Y_L=-i\lvert E_{0000,a1111}\rangle\langle E_{1111,a1111}\rvert+i\lvert E_{1111,a1111}\rangle\langle E_{0000,a1111}\rvert\). For Fig.~4(b), independent data-label amplitude damping has \(T_1=30~\mu\mathrm{s}\) and \(\eta(t)=e^{-t/T_1}\), with
\[
\begin{aligned}
\lvert1_L\rangle&=\lvert E_{1111,a1111}\rangle,\\
P_{1_L}(t)&=\langle E_{1111,a1111}|\rho(t)|E_{1111,a1111}\rangle .
\end{aligned}
\tag{A14}
\]

\subsection{Model domains}

The qutrit local model defines only the recovery subpulse. The complete local calculation uses four levels and full charge couplings for the ordered local waveform. The ring applies the static RWA of Eq.~(A6) and propagates the driven waveform in the complete \(4^8\)-dimensional product space; Fig.~4 uses the effective channel in Eq.~(A10).

\bibliography{references}
\end{document}